\documentclass[10pt,final,journal,twocolumn]{IEEEtran}
\ifCLASSINFOpdf

\else
\usepackage{algorithm}
\usepackage{booktabs}

\fi
\usepackage{cite}
\usepackage[cmex10]{amsmath}
\usepackage{amsthm}
\usepackage{stfloats}
\usepackage{multicol,multienum}
\usepackage{multirow}

\usepackage{amssymb}
\usepackage{url}
\usepackage{amsmath}
\usepackage{xcolor}
\usepackage[dvips]{graphicx}
\usepackage{subfigure}
\usepackage{caption}
\usepackage{amsmath}
\usepackage{amsfonts,amssymb}
\usepackage{bbm}
\usepackage[T1]{fontenc}
\usepackage{algorithm}
\usepackage{algpseudocode}
\usepackage{ulem}

\usepackage[linesnumbered, ruled]{}
\makeatletter
\renewcommand{\maketag@@@}[1]{\hbox{\m@th\normalsize\normalfont#1}}%
\makeatother

\begin{document}
\hyphenation{op-tical net-works semi-conduc-tor}
\vspace{-1.75em}
\title{\vspace{-0.7em} \LARGE Movable-Antenna-Array-Enabled Communications with CoMP Reception}\vspace{-1.5em}
\author{Guojie Hu, Qingqing Wu,~\textit{Senior Member}, \textit{IEEE}, Jian Ouyang,~\textit{Member}, \textit{IEEE}, Kui Xu,~\textit{Member}, \textit{IEEE}, Yunlong Cai,~\textit{Senior Member}, \textit{IEEE}, and Naofal Al-Dhahir,~\textit{Fellow}, \textit{IEEE}\vspace{-1.4em}
\thanks{
%
This work was supported in part by the Natural Science Foundations of China under Grants 62201606.
Guojie Hu is with the College of Communication Engineering, Rocket Force University of Engineering, Xi'an 710025, China (lgdxhgj@sina.com). Qingqing Wu is with the Department
of Electronic Engineering, Shanghai Jiao Tong University, Shanghai 200240, China (qingqingwu@sjtu.edu.cn). Jian Ouyang is with the Institute of Signal Processing and Transmission, Nanjing University of Posts and Telecommunications, Nanjing 210003, China (ouyangjian@njupt.edu.cn). Kui Xu is with the College of Communications Engineering, the Army of Engineering University, Nanjing 210007, China (lgdxxukui@sina.com). Yunlong Cai is with the College of Information Science and Electronic Engineering, Zhejiang University, Hangzhou 310027, China (ylcai@zju.edu.cn). Naofal Al-Dhahir is with the Department of Electrical and Computer Engineering, The University of Texas at Dallas, Richardson, TX 75080 USA (aldhahir@utdalls.edu).
}\vspace{-1.4em}
}
\IEEEpeerreviewmaketitle
\maketitle
\begin{abstract}
We consider the movable-antenna (MA) array-enabled wireless communication with coordinate multi-point (CoMP) reception, where multiple destinations adopt the maximal ratio combination technique to jointly decode the common message sent from the transmitter equipped with the MA array. Our goal is to maximize the effective received signal-to-noise ratio, by jointly optimizing the transmit beamforming and the positions of the MA array. Although the formulated problem is highly non-convex, we reveal that it is fundamental to maximize the principal eigenvalue of a hermite channel matrix which is a function of the positions of the MA array. The corresponding sub-problem is still non-convex, for which we develop a computationally efficient algorithm. Afterwards, the optimal transmit beamforming is determined with a closed-form solution. In addition, the theoretical performance upper bound is analyzed. Since the MA array brings an additional spatial degree of freedom by flexibly adjusting all antennas' positions, it achieves significant performance gain compared to competitive benchmarks.
\end{abstract}

\begin{IEEEkeywords}
Movable antenna array, coordinate multi-point reception, principal eigenvalue, positions of antennas, transmit beamforming.
\end{IEEEkeywords}

\IEEEpeerreviewmaketitle
\vspace{-15pt}
\section{Introduction}
With the ability of focusing more signal power in the desired directions, beamforming has played an important role in wireless communications. Classical transmit beamforming just pre-processes the amplitude and phase of the signal at each antenna of the transmitter to achieve signal coherent superposition at the destination(s), which, however, has its fundamental deficiency because such beamforming relies on the fixed channel conditions [1]. That is to say, if the transmission channels are unfavorable, beamforming cannot fully unleash its significant advantages.

Motivated by the above challenge, a natural idea is whether the transmission channels can be reconfigured to be most suitable for beamforming? Fortunately, the recently emerging movable antenna (MA) technology (or fluid antenna system [2]$-$[3]) may provide a feasible answer [4]. Specifically, unlike the fixed-position antenna (FPA) system, all antennas in the MA system can move in the specified region to proactively vary the steering vectors corresponding to different angles, based on which the transmission channels between transmitter and destination(s) can be flexibly adjusted. By exploiting this additional spatial degree of freedom (DoF), the MA system has shown its significant potential in enhancing the capacity of multi-input multi-output (MIMO) systems [5]$-$[6] and multi-user communications [7]$-$[10], nulling the signal power in undesired directions [11]$-$[12] or securing wireless communications [13].

Different from existing woks, this letter studies the MA array-enabled wireless communication with coordinate multi-point (CoMP) reception, where multiple destinations jointly decode the common signal sent from the transmitter equipped with the MA array. The related scenarios are practically relevant. For instance, multiple destinations may represent distributed antennas or access points (APs) in the cell-free network, which can jointly transmit the received message to a central processing unit (CPU) via their high-speed backhaul link such as X2 interface in Long Term Evolution (LTE) [14]$-$[15]. Under this setup, we aim to maximize the effective received signal-to-noise ratio (or equivalently the achievable rate) at the destinations, by jointly optimizing the transmit beamforming and positions of all antennas at the transmitter. The main contributions are summarized as follows.
 \begin{itemize}
 \item The formulated problem is highly non-convex. However, we reveal that given positions of all antennas, the optimal beamforming admits a closed-form solution, substituting which into the objective, the problem can be simplified into maximizing the principal eigenvalue of a hermite matrix, which is just related to positions of all antennas.
 \item The simplified problem is still non-convex. We develop a novel minorization maximization (MM) algorithm, which directly applies to the principal eigenvalue instead of the original objective. Thus, we propose new optimization procedures to iteratively find a locally optimal solution.
 \item Compared to the conventional alternating optimizing (AO) algorithm, our proposed algorithm provides insightful conclusion, and achieves the same performance but with much lower complexity. Also, since the MA array can flexibly change the antennas' positions to proactively cater to effective beamforming, it can achieve significant performance gain compared to competitive benchmarks.
 \end{itemize}

 \begin{figure}[t]
\centering
\includegraphics[width=6cm]{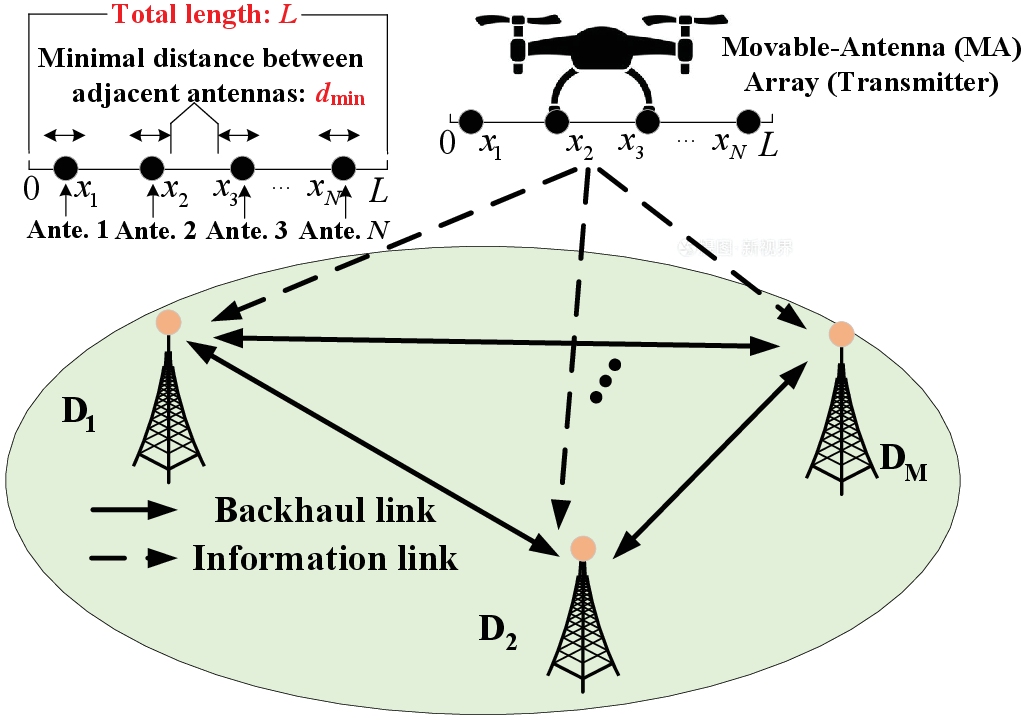}
\captionsetup{font=small}
\caption{A typical scenario of the considered system model.} \label{fig:Fig1}
\vspace{-20pt}
\end{figure}
 %


 \newcounter{mytempeqncnt}
  \begin{figure*}[b!]
  \vspace{-5pt}
  \hrulefill
\setcounter{mytempeqncnt}{\value{equation}}
\setcounter{equation}{9}
\begin{equation}
\begin{split}{}
&{\lambda _{\max }}\left( {{\bf{B}}\left( {{\bf{x}},\left\{ {{\theta _m}} \right\}_{m = 1}^M} \right)} \right) \ge {\lambda _{\max }}\left( {{\bf{B}}\left( {{{\bf{x}}^{(k)}},\left\{ {{\theta _m}} \right\}_{m = 1}^M} \right)} \right)\\
 +& \underbrace {{\rm{Tr}}\left( {{{\bf{v}}_{\max }}\left( {{\bf{B}}\left( {{{\bf{x}}^{(k)}},\left\{ {{\theta _m}} \right\}_{m = 1}^M} \right)} \right){\bf{v}}_{\max }^H\left( {{\bf{B}}\left( {{{\bf{x}}^{(k)}},\left\{ {{\theta _m}} \right\}_{m = 1}^M} \right)} \right){\bf{B}}\left( {{\bf{x}},\left\{ {{\theta _m}} \right\}_{m = 1}^M} \right)} \right)}_{{f_1}\left( {{\bf{B}}\left( {{\bf{x}},\left\{ {{\theta _m}} \right\}_{m = 1}^M} \right)} \right)}\\
 -& {\rm{Tr}}\left( {{{\bf{v}}_{\max }}\left( {{\bf{B}}\left( {{{\bf{x}}^{(k)}},\left\{ {{\theta _m}} \right\}_{m = 1}^M} \right)} \right){\bf{v}}_{\max }^H\left( {{\bf{B}}\left( {{{\bf{x}}^{(k)}},\left\{ {{\theta _m}} \right\}_{m = 1}^M} \right)} \right){\bf{B}}\left( {{{\bf{x}}^{(k)}},\left\{ {{\theta _m}} \right\}_{m = 1}^M} \right)} \right).
\end{split}
\end{equation}
\setcounter{equation}{\value{mytempeqncnt}}
\vspace{-15pt}
\end{figure*}

 \vspace{-5pt}
\section{System Model and Problem Formulation}
We consider a communication scenario as illustrated in Fig. 1, where the transmitter (S) intends to transmit a common message to $M$ destinations $\left\{ {{{\rm{D}}_m}} \right\}_{m = 1}^M$, which can cooperate in CoMP reception to jointly decode the message of S. Each of $\left\{ {{{\rm{D}}_m}} \right\}_{m = 1}^M$ is equipped with a single and fixed-position antenna, while S is equipped with a linear MA array of size $N$. By connecting the MAs to radio frequency (RF) chains via flexible cables, the MAs' positions can be flexibly adjusted in a given one-dimensional line segment of length $L$, by controller in real time, such as stepper motors or servos [4]$-$[12]. In detail, denote $x_n$ as the position of the $n$th MA, and $N$ positions are indicated by a $N \times 1$ vector ${\bf{x}} = {\left[ {{x_1},{x_2},...,{x_N}} \right]^T}$, where $0 \le {x_1} < {x_2} < ... < {x_N} \le L$ without loss of generality. Hence, given ${\bf{x}}$ and the steering angle with respect to (w.r.t.) the linear MA array as $\theta $, the steering vector of the MA array is expressed as
\setlength\abovedisplayskip{1.5pt}
\setlength\belowdisplayskip{1.5pt}
\begin{equation}
\begin{split}{}
{\bf{a}}\left( {{\bf{x}},\theta } \right) = {\left[ {{e^{j\frac{{2\pi }}{\lambda }{x_1}\cos \theta }},{e^{j\frac{{2\pi }}{\lambda }{x_2}\cos \theta }},...,{e^{j\frac{{2\pi }}{\lambda }{x_N}\cos \theta }}} \right]^T},
\end{split}
\end{equation}
where $\lambda $ is the wavelength. Denote the digital transmit beamforming of S as ${\bf{w}} \in {{\mathbb{C}}^{N \times 1}}$, with $\left\| {\bf{w}} \right\|_2^2 \le {P_S}$ and $P_S$ is the power budget of S. Then, the beam gain of the MA array at the angle $\theta $ is given by
\begin{equation}
{G_{{\bf{x}},{\bf{w}}}}(\theta ) = {\left| {{{\bf{a}}^H}\left( {{\bf{x}},\theta } \right){\bf{w}}} \right|^2},\theta  \in [0,\pi ).
\end{equation}

With CoMP reception, $M$ destinations jointly decode the received message via maximal ratio combining (MRC)\footnotemark \footnotetext{To implement MRC, each destination needs to obtain the global channel state information (CSI). This can be easily realized as follows: S first estimates the CSI by just estimating the steering angle to where each destination is located based on some mature algorithms, such as MUSIC. Then, S can inform each destination the estimated CSI. Afterwards, each destination can pro-process its received signal by multiplying the MRC coefficient and then transmit the processed signal to the CPU via the backhaul link.}, leading to the effective received signal-to-noise ratio (SNR) at $\left\{ {{{\rm{D}}_m}} \right\}_{m = 1}^M$ as
\begin{equation}
\gamma ({\bf{x}},{\bf{w}}) = \sum\nolimits_{m = 1}^M {{G_{{\bf{x}},{\bf{w}}}}({\theta _m})/{\sigma ^2}} ,
\end{equation}
where ${{\theta _m}}$ is the steering angle to where ${\rm{D}}_m$ is located and ${{\sigma ^2}}$ is the receiver noise power.

We aim to maximize $\gamma ({\bf{x}},{\bf{w}})$ (or equivalently, the achievable rate ${\log _2}(1 + \gamma ({\bf{x}},{\bf{w}}))$), by jointly optimizing the transmit beamforming ${\bf{w}}$ and the positions of $N$ MAs ${\bf{x}}$ at S. Therefore, the optimization problem is formulated as
 \begin{align}
&({\rm{P1}}):{\rm{  }}\mathop {\max }\limits_{{{\bf{x}},{\bf{w}}}} \ \gamma ({\bf{x}},{\bf{w}})\tag{${\rm{4a}}$}\\
{\rm{              }}&\ {\rm{s.t.}} \quad \ {x_n} - {x_{n - 1}} \ge {d_{\min }},n = 2,3,...,N,\tag{${\rm{4b}}$}\\
&\quad \quad \ \ \left\{ {{x_n}} \right\}_{n = 1}^N \in [0,L],\tag{${\rm{4c}}$}\\
 & \quad \ \ \quad \left\| {\bf{w}} \right\|_2^2 \le {P_S},\tag{${\rm{4d}}$}
 \end{align}
where ${d_{\min }}$ in (4b) is the minimum distance between any two adjacent MAs for avoiding the coupling effect.

Compared to the conventional FPA system, the flexible deployments of $N$ MAs will bring an additional spatial DoF for enhancing the received SNR. However, the introduction of the additional variable ${\bf{x}}$ also results in the high non-convexity of (P1), since: i) the objective of (P1) is non-concave w.r.t. ${\bf{x}}$ or ${\bf{w}}$; ii) the variables ${\bf{x}}$ and ${\bf{w}}$ are coupled with each other.
 \vspace{-4pt}
\section{Algorithm Design}
\vspace{-4pt}
In this section, an effective algorithm is developed to solve (P1). In particular, unlike most related literature, in the proposed algorithm, it is not necessary to alternately optimize ${\bf{x}}$ or ${\bf{w}}$ which causes a higher computational complexity. Instead, the optimal ${\bf{x}}$ can be first determined, based on which the optimal ${\bf{w}}$ can then be given accordingly. The details are shown as follows.

We start by expanding $\gamma ({\bf{x}},{\bf{w}})$ in (3) as
\setlength\abovedisplayskip{1.5pt}
\setlength\belowdisplayskip{1.5pt}
\begin{equation}
\setcounter{equation}{5}
\begin{split}{}
&\gamma ({\bf{x}},{\bf{w}}) = \sum\nolimits_{m = 1}^M {{{\left| {{{\bf{a}}^H}\left( {{\bf{x}},{\theta _m}} \right){\bf{w}}} \right|}^2}/{\sigma ^2}} \\
 =& \sum\nolimits_{m = 1}^M {{\rm{Tr}}\left( {{\bf{a}}\left( {{\bf{x}},{\theta _m}} \right){{\bf{a}}^H}\left( {{\bf{x}},{\theta _m}} \right){\bf{W}}} \right)} \\
 =& {\rm{Tr}}\left( {{\bf{A}}\left( {{\bf{x}},\left\{ {{\theta _m}} \right\}_{m = 1}^M} \right){{\bf{A}}^H}\left( {{\bf{x}},\left\{ {{\theta _m}} \right\}_{m = 1}^M} \right){\bf{W}}} \right),
\end{split}
\end{equation}
where ${\bf{W}} = {\bf{w}}{{\bf{w}}^H}/{\sigma ^2} \in {{\mathbb{C}}^{N \times N}}$ and ${\bf{A}}\left( {{\bf{x}},\left\{ {{\theta _m}} \right\}_{m = 1}^M} \right) = \left[ {{\bf{a}}\left( {{\bf{x}},{\theta _1}} \right),{\bf{a}}\left( {{\bf{x}},{\theta _2}} \right),...,{\bf{a}}\left( {{\bf{x}},{\theta _M}} \right)} \right] \in {{\mathbb{C}}^{N \times M}}$.

Based on (5), for any given ${\bf{x}}$, the optimal ${\bf{W}}$ admits a closed-form solution as [16]
\begin{equation}
\begin{split}{}
{{\bf{W}}^*} =& {P_S}{{\bf{v}}_{\max }}\left( {{\bf{B}}\left( {{\bf{x}},\left\{ {{\theta _m}} \right\}_{m = 1}^M} \right)} \right)  \\
&\times {\bf{v}}_{\max }^H\left( {{\bf{B}}\left( {{\bf{x}},\left\{ {{\theta _m}} \right\}_{m = 1}^M} \right)} \right)/{\sigma ^2},
\end{split}
\end{equation}
where ${{\bf{v}}_{\max }}\left( {{\bf{B}}\left( {{\bf{x}},\left\{ {{\theta _m}} \right\}_{m = 1}^M} \right)} \right) \in {{\mathbb{C}}^{N \times 1}}$ is the principal eigenvector of ${\bf{B}}\left( {{\bf{x}},\left\{ {{\theta _m}} \right\}_{m = 1}^M} \right) \buildrel \Delta \over =  {\bf{A}}\left( {{\bf{x}},\left\{ {{\theta _m}} \right\}_{m = 1}^M} \right){{\bf{A}}^H}\left( {{\bf{x}},\left\{ {{\theta _m}} \right\}_{m = 1}^M} \right) \in {{\mathbb{C}}^{N \times N}}$, with
\begin{equation}
{{\bf{B}}_{i,j}}\left( {{\bf{x}},\left\{ {{\theta _m}} \right\}_{m = 1}^M} \right) = \sum\nolimits_{m = 1}^M {{e^{j\frac{{2\pi }}{\lambda }({x_i} - {x_j})\cos {\theta _m}}}},
\end{equation}
and ${{\bf{B}}_{i,j}}\left( {{\bf{x}},\left\{ {{\theta _m}} \right\}_{m = 1}^M} \right)$ is the element in the $i$th row and $j$th column of ${\bf{B}}\left( {{\bf{x}},\left\{ {{\theta _m}} \right\}_{m = 1}^M} \right)$ according to (1).

 Substituting ${{\bf{W}}^*}$ into (5), the received SNR can be simplified as
\begin{equation}
\gamma ({\bf{x}}) = {P_S}{\lambda _{\max }}\left( {{\bf{B}}\left( {{\bf{x}},\left\{ {{\theta _m}} \right\}_{m = 1}^M} \right)} \right)/{\sigma ^2},
\end{equation}
where ${\lambda _{\max }}\left( {{\bf{B}}\left( {{\bf{x}},\left\{ {{\theta _m}} \right\}_{m = 1}^M} \right)} \right)$ is the principal eigenvalue of ${{\bf{B}}\left( {{\bf{x}},\left\{ {{\theta _m}} \right\}_{m = 1}^M} \right)}$.

Based on (8), to maximize $\gamma ({\bf{x}},{\bf{w}})$, it is fundamental to maximize ${\lambda _{\max }}\left( {{\bf{B}}\left( {{\bf{x}},\left\{ {{\theta _m}} \right\}_{m = 1}^M} \right)} \right)$, which is just related to ${\bf{x}}$. In other words, once ${\lambda _{\max }}\left( {{\bf{B}}\left( {{\bf{x}},\left\{ {{\theta _m}} \right\}_{m = 1}^M} \right)} \right)$ is maximized, ${{\bf{W}}^*}$ can then be determined based on the known ${{\bf{B}}\left( {{\bf{x}},\left\{ {{\theta _m}} \right\}_{m = 1}^M} \right)}$. Therefore, the optimization problem becomes
\setlength\abovedisplayskip{1.5pt}
\setlength\belowdisplayskip{1.5pt}
 \begin{align}
&({\rm{P2}}):{\rm{  }}\mathop {\max }\limits_{{{\bf{x}}}} \ {\lambda _{\max }}\left( {{\bf{B}}\left( {{\bf{x}},\left\{ {{\theta _m}} \right\}_{m = 1}^M} \right)} \right)\tag{${\rm{9a}}$}\\
{\rm{              }}&\ {\rm{s.t.}} \quad \ (4{\rm{b}}), (4{\rm{c}}).\tag{${\rm{9b}}$}
 \end{align}

    \begin{figure*}[b!]
  \hrulefill
\setcounter{mytempeqncnt}{\value{equation}}
\setcounter{equation}{11}
\begin{equation}
\begin{split}{}
&{f_1}\left( {{\bf{B}}\left( {{\bf{x}},\left\{ {{\theta _m}} \right\}_{m = 1}^M} \right)} \right) = \sum\nolimits_{i = 1}^N {\sum\nolimits_{j = 1}^N {{{\bf{v}}_{\max ,i}}\left( {{\bf{B}}\left( {{{\bf{x}}^{(k)}},\left\{ {{\theta _m}} \right\}_{m = 1}^M} \right)} \right){\bf{v}}_{\max ,j}^H\left( {{\bf{B}}\left( {{{\bf{x}}^{(k)}},\left\{ {{\theta _m}} \right\}_{m = 1}^M} \right)} \right)} } \\
&\times {{\bf{B}}_{j,i}}\left( {{\bf{x}},\left\{ {{\theta _m}} \right\}_{m = 1}^M} \right)\mathop  = \limits^{(a)} M\sum\nolimits_{i = 1}^N {{{\left| {{{\bf{v}}_{\max ,i}}\left( {{\bf{B}}\left( {{{\bf{x}}^{(k)}},\left\{ {{\theta _m}} \right\}_{m = 1}^M} \right)} \right)} \right|}^2}} \\
 &+ 2\underbrace {{\mathop{\rm Re}\nolimits} \left[ {\sum\nolimits_{i = 1}^{N - 1} {\sum\nolimits_{j = i + 1}^N {{{\bf{v}}_{\max ,i}}\left( {{\bf{B}}\left( {{{\bf{x}}^{(k)}},\left\{ {{\theta _m}} \right\}_{m = 1}^M} \right)} \right){\bf{v}}_{\max ,j}^H\left( {{\bf{B}}\left( {{{\bf{x}}^{(k)}},\left\{ {{\theta _m}} \right\}_{m = 1}^M} \right)} \right){{\bf{B}}_{j,i}}} } \left( {{\bf{x}},\left\{ {{\theta _m}} \right\}_{m = 1}^M} \right)} \right]}_{{f_2}\left( {{\bf{B}}\left( {{\bf{x}},\left\{ {{\theta _m}} \right\}_{m = 1}^M} \right)} \right)}.
\end{split}
\end{equation}
\setcounter{equation}{\value{mytempeqncnt}}
\vspace{-15pt}
\end{figure*}

Although problem (P2) is still highly non-convex, the objective, i.e., ${\lambda _{\max }}\left( {{\bf{B}}\left( {{\bf{x}},\left\{ {{\theta _m}} \right\}_{m = 1}^M} \right)} \right)$, is convex w.r.t. ${{\bf{B}}\left( {{\bf{x}},\left\{ {{\theta _m}} \right\}_{m = 1}^M} \right)}$. Note that the convex function can be globally lower-bounded by its first-order Taylor expansion. Hence, with given ${{\bf{x}}^{(k)}}$ in the $k$th iteration, ${\lambda _{\max }}\left( {{\bf{B}}\left( {{\bf{x}},\left\{ {{\theta _m}} \right\}_{m = 1}^M} \right)} \right)$ has a lower bound presented in (10).

 By neglecting the irrelevant constant components, the optimization problem in the $k$th iteration can be relaxed as
 \begin{align}
&({\rm{P2.1}}):{\rm{  }}\mathop {\max }\limits_{{{\bf{x}}}} \ {f_1}\left( {{\bf{B}}\left( {{\bf{x}},\left\{ {{\theta _m}} \right\}_{m = 1}^M} \right)} \right)\tag{${\rm{11a}}$}\\
{\rm{              }}&\ {\rm{s.t.}} \quad \ (4{\rm{b}}), (4{\rm{c}}),\tag{${\rm{11b}}$}
 \end{align}
where the objective ${{f_1}\left( {{\bf{B}}\left( {{\bf{x}},\left\{ {{\theta _m}} \right\}_{m = 1}^M} \right)} \right)}$ can be expanded as in (12), in which ${{{\bf{v}}_{\max ,i}}\left( {{\bf{B}}\left( {{{\bf{x}}^{(k)}},\left\{ {{\theta _m}} \right\}_{m = 1}^M} \right)} \right)}$ is the element in the $i$th row of ${{{\bf{v}}_{\max }}\left( {{\bf{B}}\left( {{{\bf{x}}^{(k)}},\left\{ {{\theta _m}} \right\}_{m = 1}^M} \right)} \right)}$, and $\mathop  = \limits^{(a)} $ is established because: i) ${{\bf{B}}_{i,i}}\left( {{\bf{x}},\left\{ {{\theta _m}} \right\}_{m = 1}^M} \right) = M$ for any $i = 1,2,...,N$; ii) ${{\bf{v}}_{\max ,i}}{\bf{v}}_{\max ,j}^H{{\bf{B}}_{j,i}} = {\left[ {{{\bf{v}}_{\max ,j}}{\bf{v}}_{\max ,i}^H{{\bf{B}}_{i,j}}} \right]^H}$ (since ${\bf{B}}\left( {{\bf{x}},\left\{ {{\theta _m}} \right\}_{m = 1}^M} \right)$ is a hermite matrix) and then ${{\bf{v}}_{\max ,i}}{\bf{v}}_{\max ,j}^H{{\bf{B}}_{j,i}} + {{\bf{v}}_{\max ,j}}{\bf{v}}_{\max ,i}^H{{\bf{B}}_{i,j}} = 2{\mathop{\rm Re}\nolimits} \left[ {{{\bf{v}}_{\max ,i}}{\bf{v}}_{\max ,j}^H{{\bf{B}}_{j,i}}} \right]$ for any $i \ne j$.

 By neglecting the irrelevant constant components and based on the expression of ${{\bf{B}}_{i,j}}\left( {{\bf{x}},\left\{ {{\theta _m}} \right\}_{m = 1}^M} \right)$ shown in (7), (P2.1) can be further simplified as
  \begin{align}
&({\rm{P2.2}}):{\rm{  }}\mathop {\max }\limits_{{{\bf{x}}}} \ {f_2}\left( {{\bf{B}}\left( {{\bf{x}},\left\{ {{\theta _m}} \right\}_{m = 1}^M} \right)} \right)\tag{${\rm{13a}}$}\\
 =& {\mathop{\rm Re}\nolimits} \left[ {\sum\nolimits_{i = 1}^{N - 1} {\sum\nolimits_{j = i + 1}^N {\sum\nolimits_{m = 1}^M {{o_i}o_j^H{e^{j\frac{{2\pi }}{\lambda }({x_j} - {x_i})\cos {\theta _m}}}} } } } \right]\nonumber \\
 =& \sum\nolimits_{i = 1}^{N - 1} {\sum\nolimits_{j = i + 1}^N {\sum\nolimits_{m = 1}^M {\left| {{o_i}} \right|\left| {o_j^H} \right|} } }\nonumber \\
 &\times \cos \left( {{\varphi _{{o_i}}} + {\varphi _{o_j^H}} + \frac{{2\pi ({x_j} - {x_i})\cos {\theta _m}}}{\lambda }} \right)\nonumber\\
{\rm{              }}&\ {\rm{s.t.}} \quad \ (4{\rm{b}}), (4{\rm{c}}),\tag{${\rm{13b}}$}
 \end{align}
 where ${o_i} = {{\bf{v}}_{\max ,i}}\left( {{\bf{B}}\left( {{{\bf{x}}^{(k)}},\left\{ {{\theta _m}} \right\}_{m = 1}^M} \right)} \right)$, $o_j^H = {\bf{v}}_{\max ,j}^H\left( {{\bf{B}}\left( {{{\bf{x}}^{(k)}},\left\{ {{\theta _m}} \right\}_{m = 1}^M} \right)} \right)$, ${{\varphi _{{o_i}}}}$ and ${{\varphi _{o_j^H}}}$ are the phases of ${{o_i}}$ and ${o_j^H}$, respectively.

 Unfortunately, ${f_2}\left( {{\bf{B}}\left( {{\bf{x}},\left\{ {{\theta _m}} \right\}_{m = 1}^M} \right)} \right)$ is still non-convex or non-concave w.r.t. ${\bf{x}}$ because of the $\cos ( \cdot )$ function. Nevertheless, based on the minorization maximization (MM) algorithm [17], we can construct a surrogate function that locally approximates the objective by exploiting the second-order Taylor expansion. Specifically, according to Taylor's theorem, given ${{{\bf{x}}^{(k)}}}$ in the $k$th iteration, a quadratic surrogate function which serves the lower bound of the objective ${f_2}\left( {{\bf{B}}\left( {{\bf{x}},\left\{ {{\theta _m}} \right\}_{m = 1}^M} \right)} \right)$ is derived as [17]
 \begin{equation}
\setcounter{equation}{14}
\begin{split}{}
&{f_2}\left( {{\bf{B}}\left( {{\bf{x}},\left\{ {{\theta _m}} \right\}_{m = 1}^M} \right)} \right) \ge {f_2}\left( {{\bf{B}}\left( {{{\bf{x}}^{(k)}},\left\{ {{\theta _m}} \right\}_{m = 1}^M} \right)} \right)\\
 &+ \nabla {f_2}{\left( {{\bf{B}}\left( {{{\bf{x}}^{(k)}},\left\{ {{\theta _m}} \right\}_{m = 1}^M} \right)} \right)^T}({\bf{x}} - {{\bf{x}}^{(k)}})\\
 &- \frac{\delta }{2}{({\bf{x}} - {{\bf{x}}^{(k)}})^T}({\bf{x}} - {{\bf{x}}^{(k)}})\buildrel \Delta \over = {f_3}\left( {{\bf{B}}\left( {{{\bf{x}}^{(k)}},\left\{ {{\theta _m}} \right\}_{m = 1}^M} \right),{\bf{x}}} \right)
\end{split}
\end{equation}
where $\nabla {f_2}\left( {{\bf{B}}\left( {{{\bf{x}}^{(k)}},\left\{ {{\theta _m}} \right\}_{m = 1}^M} \right)} \right) \in {{\mathbb{R}}^{N \times 1}}$ is the gradient vector of ${f_2}\left( {{\bf{B}}\left( {{\bf{x}},\left\{ {{\theta _m}} \right\}_{m = 1}^M} \right)} \right)$ at ${{{\bf{x}}^{(k)}}}$, the positive real number $\delta $ should satisfy $\delta {{\bf{I}}_N}\underline  \succ  {\nabla ^2}{f_2}\left( {{\bf{B}}\left( {{\bf{x}},\left\{ {{\theta _m}} \right\}_{m = 1}^M} \right)} \right)$, where ${\nabla ^2}{f_2}\left( {{\bf{B}}\left( {{\bf{x}},\left\{ {{\theta _m}} \right\}_{m = 1}^M} \right)} \right) \in {{\mathbb{R}}^{N \times N}}$ is the Hessian matrix of ${f_2}\left( {{\bf{B}}\left( {{\bf{x}},\left\{ {{\theta _m}} \right\}_{m = 1}^M} \right)} \right)$. The expressions of $\nabla {f_2}\left( {{\bf{B}}\left( {{{\bf{x}}^{(k)}},\left\{ {{\theta _m}} \right\}_{m = 1}^M} \right)} \right)$ and $\delta $ are given in Appendix A.

Based on (14), (P2.2) can be finally relaxed as
 \begin{align}
&({\rm{P2.3}}):{\rm{  }}\mathop {\max }\limits_{{{\bf{x}}}} \ {f_3}\left( {{\bf{B}}\left( {{{\bf{x}}^{(k)}},\left\{ {{\theta _m}} \right\}_{m = 1}^M} \right),{\bf{x}}} \right)\tag{${\rm{15a}}$}\\
{\rm{              }}&\ {\rm{s.t.}} \quad \ (4{\rm{b}}), (4{\rm{c}}),\tag{${\rm{15b}}$}
 \end{align}
which is a convex problem and thus can be solved using standard convex optimization techniques such as interior point method or CVX. The overall procedures for solving (P1) are presented in Algorithm 1.

\begin{figure*}[t]
\vspace{-12pt}
\centering

\begin{minipage}{5.5cm}
\includegraphics[width=5.5cm]{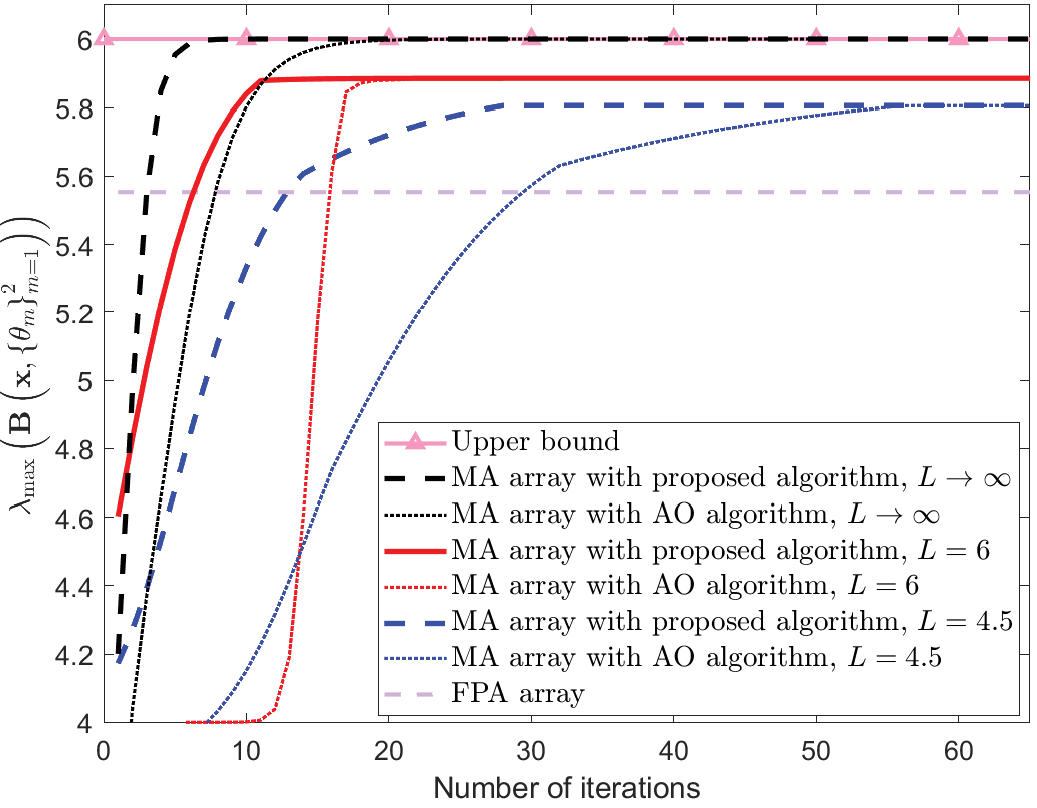}
\centering
\subfigure{(a)}

\end{minipage}
\begin{minipage}{5.15cm}
\includegraphics[width=5.15cm]{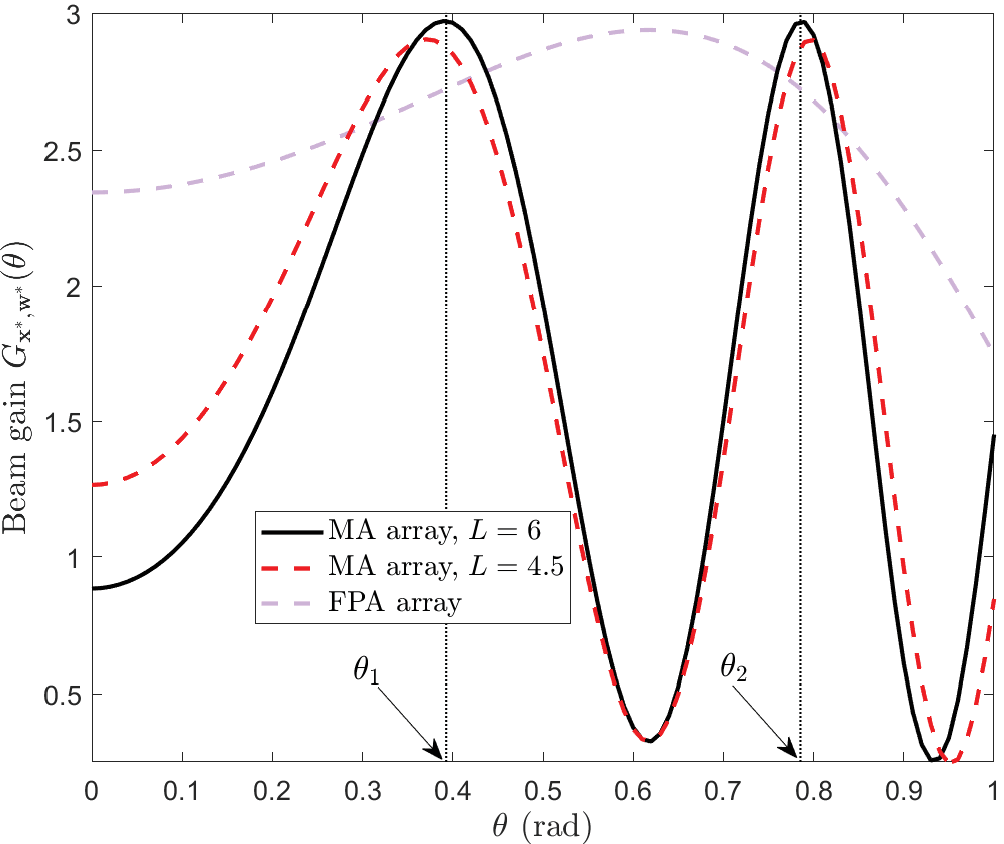}
\centering
\subfigure{(b)}

\end{minipage}
\begin{minipage}{5.5cm}
\includegraphics[width=5.5cm]{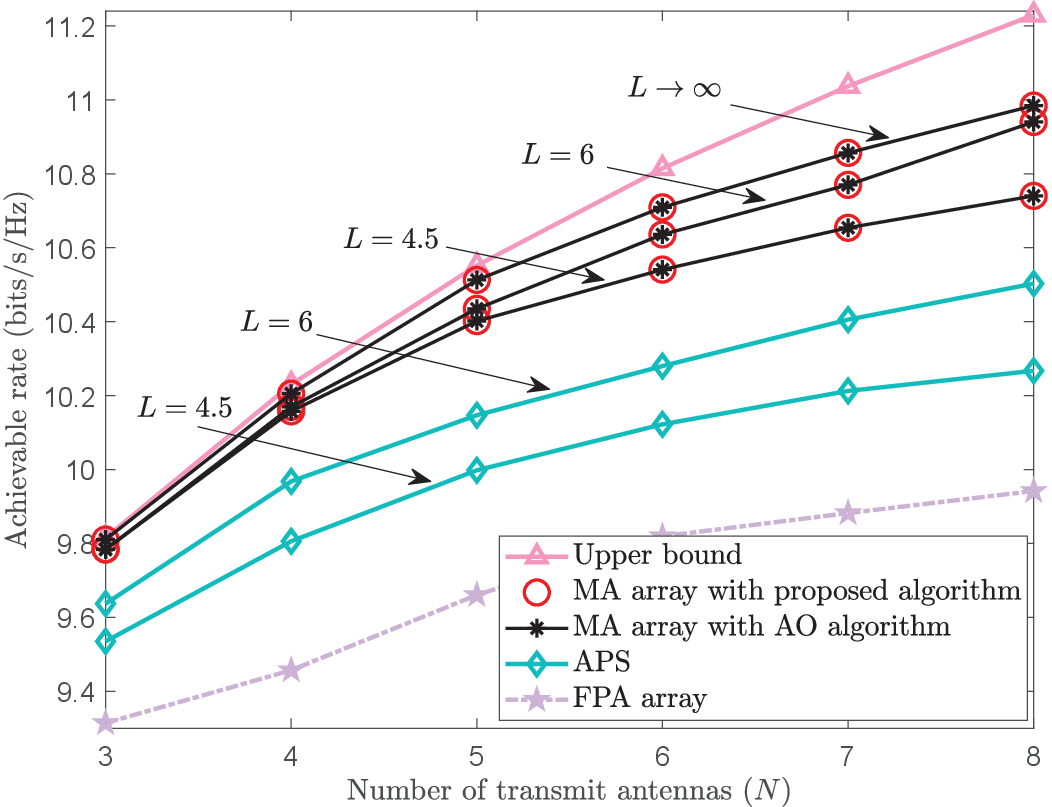}
\centering
\subfigure{(c)}

\end{minipage}
\caption{(a) Convergence behavior of the proposed algorithm and the AO algorithm; (b) Beam gain of the MA array and the FPA array; (c) Achievable rate w.r.t. the number of transmit antennas at S ($N$).}
\vspace{-20pt}
\end{figure*}

\begin{algorithm}
\caption{Proposed MM Algorithm for Solving (P1)}
  \begin{algorithmic}[1]
\State Initialize the MA array positions as ${{\bf{x}}^{(0)}} = {\left[ {0,L/(N - 1),2L/(N - 1),...,L} \right]^T}$ and $k = 0$.

\State \textbf{Repeat:}

\State \quad Compute ${{\bf{v}}_{\max }}\left( {{\bf{B}}\left( {{{\bf{x}}^{(k)}},\left\{ {{\theta _m}} \right\}_{m = 1}^M} \right)} \right)$, based on which

\noindent \quad calculate $\nabla {f_2}\left( {{\bf{B}}\left( {{{\bf{x}}^{(k)}},\left\{ {{\theta _m}} \right\}_{m = 1}^M} \right)} \right)$ and $\delta $;

\State \quad Solve (P2.3) and obtain the optimized ${{\bf{x}}^*}$;

\State \quad $k \leftarrow k + 1$, ${{\bf{x}}^{(k)}} \leftarrow {{\bf{x}}^*}$;


\State \textbf{Until:} The objective of (P2) converges to a prescribed accuracy.

\State Compute ${{\bf{v}}_{\max }}\left( {{\bf{B}}\left( {{{\bf{x}}^*},\left\{ {{\theta _m}} \right\}_{m = 1}^M} \right)} \right)$ and then obtain ${{\bf{W}}^*}$ based on (6).
  \end{algorithmic}
\end{algorithm}

\textit{Complexity Analysis:} For Algorithm 1, in setup 3, the complexities of computing ${{\bf{v}}_{\max }}\left( {{\bf{B}}\left( {{{\bf{x}}^{(k)}},\left\{ {{\theta _m}} \right\}_{m = 1}^M} \right)} \right)$, $\nabla {f_2}\left( {{\bf{B}}\left( {{{\bf{x}}^{(k)}},\left\{ {{\theta _m}} \right\}_{m = 1}^M} \right)} \right)$ and $\delta $ are ${\cal O}\left( {{N^3}} \right)$, ${\cal O}\left( {M(N - 1)} \right)$ and ${\cal O}\left( 1 \right)$, respectively. In setup 4, solving ${{\bf{x}}^*}$ requires the complexity of ${\cal O}\left( {{N^{3.5}}\ln (1/\varepsilon )} \right)$ with accuracy $\varepsilon $. Hence, the complexity from setups 2 to 6 is about ${\cal O}\left( {KM(N - 1) + K{N^{3.5}}\ln (1/\varepsilon )} \right)$, where $K$ is the number of iterations. Further, the complexity of computing ${{\bf{v}}_{\max }}\left( {{\bf{B}}\left( {{{\bf{x}}^*},\left\{ {{\theta _m}} \right\}_{m = 1}^M} \right)} \right)$ in setup 7 is about ${\cal O}\left( {{N^3}} \right)$, leading to the total complexity of Algorithm 1 as ${\cal O}\left( {KM(N - 1) + K{N^{3.5}}\ln (1/\varepsilon )} \right)$. In addition, the output values of (P2.3) are non-decreasing and the optimal solution of (P1) has a upper bound as specified below. Hence, the algorithm convergence can be guaranteed [5].

\textit{Performance Bound:} Because the principal eigenvalue of a hermite matrix is bounded by the maximum value of the sum of the column elements of such matrix, based on the expression of ${{\bf{B}}_{i,j}}\left( {{\bf{x}},\left\{ {{\theta _m}} \right\}_{m = 1}^M} \right)$ shown in (7), it can be derived that
\setlength\abovedisplayskip{1.5pt}
\setlength\belowdisplayskip{1.5pt}
 \begin{equation}
 \setcounter{equation}{16}
\begin{split}{}
&\max \left( {{\lambda _{\max }}\left( {{\bf{B}}\left( {{\bf{x}},\left\{ {{\theta _m}} \right\}_{m = 1}^M} \right)} \right)} \right)\\
 \le& M + (N - 1)\max \left| {\sum\nolimits_{m = 1}^M {{e^{j\frac{{2\pi }}{\lambda }({x_i} - {x_j})\cos {\theta _m}}}} } \right|\\
 =& MN.
\end{split}
\end{equation}


\textbf{Remark:} As a comparison, if the AO algorithm is exploited to frequently update ${\bf{w}}$ and ${\bf{x}}$, i) the complexity of optimizing ${\bf{w}}$ given ${\bf{x}}$ is still ${\cal O}({N^3})$; ii) the complexity of optimizing ${\bf{x}}$ given ${\bf{w}}$ has been reported as ${\cal O}\left( {{{K_{{\rm{inner}}}}}M(N - 1) + {{{K_{\rm{inner}}}}}{N^{3.5}}\ln (1/\varepsilon )} \right)$ [5] since the corresponding problem becomes
  \begin{equation}
\mathop {\max }\limits_{\bf{x}} \sum\nolimits_{m = 1}^M {{{\bf{a}}^H}\left( {{\bf{x}},{\theta _m}} \right){\bf{Wa}}\left( {{\bf{x}},{\theta _m}} \right)} ,\ \ {\rm{s.t}}.,(4{\rm{b}}),(4{\rm{c}}),
\end{equation}
 with ${\bf{W}} = {\bf{w}}{{\bf{w}}^H}/{\sigma ^2}$, which has the same form as equation (25) of [5]. Thus, the total complexity would be ${{\cal O}\left( {{K_{{\rm{outer}}}}\left( {{K_{{\rm{inner}}}}M(N - 1) + {K_{{\rm{inner}}}}{N^{3.5}}\ln (1/\varepsilon )} \right)} \right)}$, where $K_{{\rm{outer}}}$ is the number of outer iteration. Obviously, such complexity is higher than that of our proposed algorithm. Moreover, different from our proposed algorithm, there is no any insightful conclusion that can be obtained from the alternating optimizing.

\vspace{-5pt}
\section{Simulation Results}
In this section, we present numerical results to validate the effectiveness of the designed scheme with the MA array. The minimum distance between any two MAs is set as ${d_{\min }} = \lambda /2$ and $\lambda$ is normalized as 1. Further, the receiver noise power is ${\sigma ^2} = 1$ for normalizing the large-scale fading power.

We first illustrate the convergence behavior of the proposed Algorithm 1 for the MA array in Fig. 2(a). The parameters are: $N = 3$, $M = 2$, ${\theta _1} = \pi /8$ and ${\theta _2} = \pi /4$. From Fig. 2(a) it is observed that: i) the objective, i.e., ${\lambda _{\max }}\left( {{\bf{B}}\left( {{\bf{x}},\left\{ {{\theta _m}} \right\}_{m = 1}^2} \right)} \right)$, converges to a constant after dozens of iterations, implying that the proposed algorithm is computationally efficient. As a comparison, the number of iterations using the AO algorithm is larger, the key reason lies in that unlike our algorithm where only ${\bf{x}}$ is optimized iteratively, ${\bf{w}}$ and ${\bf{x}}$ in the AO algorithm are optimized alternatively; ii) when the length of the line segment $L$ becomes larger, all optimization variables in ${\bf{x}}$ can be adjusted more flexibly, hence a better performance can be achieved. When $L \to \infty $, it is clear to see that ${\lambda _{\max }}\left( {{\bf{B}}\left( {{\bf{x}},\left\{ {{\theta _m}} \right\}_{m = 1}^2} \right)} \right)$ converges to its upper bound, i.e., $MN = 2 \times 3 = 6$; iii) in comparison, for the FPA array where the positions of $N$ antennas are fixed as ${{\bf{x}}^{{\rm{FPA}}}} = {[0,{d_{\min }},...,(N - 1){d_{\min }}]^T}$ and S adopts the transmit beamforming in (6), since no additional spatial DoF can be exploited, the resulting performance is unfavorable.

In Fig. 2(b), we further present the corresponding beam gain of the MA array and the FPA array at the angle $\theta $, with $\theta  \in [0,1]$ (rad). The parameters are the same as in Fig. 2(a) and $P_S = 0$ dB. We observe that: i) with the careful design for the antenna positions, the MA array can produce a higher beam gain at each of two desirable angles ${\theta _1}$ and ${\theta _2}$ compared to the FPA array. Then, the resulting received SNR of the MA array, which is proportional to the sum of ${G_{{{\bf{x}}^*},{{\bf{w}}^*}}}({\theta _1})$ and ${G_{{{\bf{x}}^*},{{\bf{w}}^*}}}({\theta _2})$, obviously is larger; ii) as $L$ increases, the beam gain at the desirable angle becomes larger, the reasons have been mentioned above and thus are omitted here.

Fig. 2(c) shows the achievable rate w.r.t. the number of transmit antennas at S ($N$), where $P_S = 20$ dB, $M = 3$, ${\theta _1} = \pi /8$, ${\theta _2} = \pi /4$ and ${\theta _3} = 3\pi /4$. For comprehensive comparison, we further consider another representative scheme called alternating position selection (APS) [5], [12], in which the line segment of length $L$ is quantized into $2L + 1$ discrete locations with equal-distance ${d_{\min }} = 1 /2$. Then, $N$ out of these $2L + 1$ locations are optimally selected for antenna positions, and S also adopts the transmit beamforming in (6). From Fig. 2(c), we observe that: i) since more transmit antennas can produce a larger spatial diversity and multiplexing, obviously the achievable rate of all schemes increases w.r.t. $N$; ii) since both of APS and the FPA array (or other schemes mentioned in [5], [12]) are the special cases of the MA array, in which all antennas can be deployed in a most flexible manner, obviously the MA array can enjoy the maximum advantage and then achieve the favorable performance; iii) the performance upper bound in (16) actually is derived under absolute ideal conditions where no any correlation exists between any elements of ${{\bf{B}}\left( {{\bf{x}},\left\{ {{\theta _m}} \right\}_{m = 1}^M} \right)}$. However, in practical situation, although positions of all antennas (${\bf{x}}$) can be flexibly adjusted, there exists the inherent correlation between certain elements of ${{\bf{B}}\left( {{\bf{x}},\left\{ {{\theta _m}} \right\}_{m = 1}^M} \right)}$. For instance, once the elements in the $i$th row and $j$th column, and in the $i$th row and $k$th column, are fixed as $\sum\nolimits_{m = 1}^M {{e^{j\frac{{2\pi }}{\lambda }({x_i} - {x_j})\cos {\theta _m}}}} $ and $\sum\nolimits_{m = 1}^M {{e^{j\frac{{2\pi }}{\lambda }({x_i} - {x_k})\cos {\theta _m}}}} $, $i \ne j \ne k$, the element in the $j$th row and $k$th column cannot be flexibly varied anymore but just equals $\sum\nolimits_{m = 1}^M {{e^{j\frac{{2\pi }}{\lambda }({x_j} - {x_k})\cos {\theta _m}}}} $. When $N$ is small, e.g., 3$-$5 in Fig. 2(c), there is a little correlation between the elements of ${{\bf{B}}\left( {{\bf{x}},\left\{ {{\theta _m}} \right\}_{m = 1}^M} \right)}$, which provides a more flexible space (DoF) for the matrix optimization. Thus, the performance of the proposed algorithm almost approaches the bound. While when $N$ becomes large, more elements of ${\bf{B}}\left( {{\bf{x}},\left\{ {{\theta _m}} \right\}_{m = 1}^M} \right)$ are correlated with each other, this inevitably reduces the DoF to maximize the principal eigenvalue. Then, there will exist certain rate gap between the proposed algorithm and the bound; iv) the performance of our proposed algorithm and the AO algorithm is almost the same, while the former owns much lower complexity, indicating that our proposed algorithm may be practically efficient.

\vspace{-20pt}
\section{Conclusion}
\vspace{-9pt}
This letter investigates the MA array-enabled wireless communication with CoMP reception. We aim to maximize the effective received SNR by optimizing the transmit beamforming and the positions of the MA array at the transmitter. The essence is to maximize the principal eigenvalue of a hermite matrix which is related to the positions of the MA array, for solving which an iterative algorithm is developed. Then, the optimal transmit beamforming is determined accordingly. It is shown by simulations the effectiveness of the proposed design with the MA array compared to competitive benchmarks.
\vspace{-15pt}
  \begin{appendices}
\section{The expressions of $\nabla {f_2}\left( {{\bf{B}}\left( {{{\bf{x}}^{(k)}},\left\{ {{\theta _m}} \right\}_{m = 1}^M} \right)} \right)$ and $\delta $}
Based on (13a), it is derived that
\begin{equation}
\setcounter{equation}{18}
\begin{split}
&\frac{{\partial {f_2}\left( {{\bf{B}}\left( {{\bf{x}},\left\{ {{\theta _m}} \right\}_{m = 1}^M} \right)} \right)}}{{\partial {x_i}}} = \sum\nolimits_{m = 1}^M {\sum\nolimits_{j = i + 1}^N {\left| {{o_i}} \right|\left| {o_j^H} \right|} } \\
\times& \sin \left( {{\varphi _{{o_i}}} + {\varphi _{o_j^H}} + \frac{{2\pi ({x_j} - {x_i})\cos {\theta _m}}}{\lambda }} \right)\frac{{2\pi \cos {\theta _m}}}{\lambda }\\
 &- \sum\nolimits_{m = 1}^M {\sum\nolimits_{j = 1}^{i - 1} {\left| {{o_j}} \right|\left| {o_i^H} \right|} } \\
\times& \sin \left( {{\varphi _{{o_j}}} + {\varphi _{o_i^H}} + \frac{{2\pi ({x_i} - {x_j})\cos {\theta _m}}}{\lambda }} \right)\frac{{2\pi \cos {\theta _m}}}{\lambda },
\end{split}
\end{equation}
  according to which it is easy to derive
 \begin{equation}
\begin{split}
&\nabla {f_2}\left( {{\bf{B}}\left( {{{\bf{x}}^{(k)}},\left\{ {{\theta _m}} \right\}_{m = 1}^M} \right)} \right)\\
 =& \left( {\left[ {\frac{{\partial {f_2}\left( {{\bf{B}}\left( {{\bf{x}},\left\{ {{\theta _m}} \right\}_{m = 1}^M} \right)} \right)}}{{\partial {x_i}}}} \right]_{i = 1}^N} \right)_{{\bf{x}} = {{\bf{x}}^{(k)}}}^T.
\end{split}
\end{equation}

On the other hand, note that
 \begin{equation}
\begin{split}
&\nabla f_2^2\left( {{\bf{B}}\left( {{\bf{x}},\left\{ {{\theta _m}} \right\}_{m = 1}^M} \right)} \right)\\
 =& {\left[ {\frac{{\partial f_2^2\left( {{\bf{B}}\left( {{\bf{x}},\left\{ {{\theta _m}} \right\}_{m = 1}^M} \right)} \right)}}{{\partial {x_i}\partial {x_j}}}} \right]_{i,j = 1,...,N}},
\end{split}
\end{equation}
and based on (18), it is determined that
 \begin{equation}
\begin{split}
\frac{{\partial f_2^2\left( {{\bf{B}}\left( {{\bf{x}},\left\{ {{\theta _m}} \right\}_{m = 1}^M} \right)} \right)}}{{\partial {x_i}\partial {x_i}}} \le& \sum\nolimits_{j = 1,j \ne i}^N {\left| {{o_i}} \right|\left| {{o_j}} \right|} Q,\\
\frac{{\partial f_2^2\left( {{\bf{B}}\left( {{\bf{x}},\left\{ {{\theta _m}} \right\}_{m = 1}^M} \right)} \right)}}{{\partial {x_i}\partial {x_j}}} \le& \left| {{o_i}} \right|\left| {{o_j}} \right|Q,
\end{split}
\end{equation}
where $i,j = 1,...,N$, $i \ne j$ and $Q = \sum\nolimits_{m = 1}^M {\frac{{4{\pi ^2}{{\cos }^2}{\theta _m}}}{{{\lambda ^2}}}} $. Then,
 \begin{equation}
\begin{split}
&\left\| {\nabla f_2^2\left( {{\bf{B}}\left( {{\bf{x}},\left\{ {{\theta _m}} \right\}_{m = 1}^M} \right)} \right)} \right\|_2^2\\
 \le& \sum\nolimits_{i = 1}^N {\sum\nolimits_{j = 1}^N {{{\left( {\frac{{\partial f_2^2\left( {{\bf{B}}\left( {{\bf{x}},\left\{ {{\theta _m}} \right\}_{m = 1}^M} \right)} \right)}}{{\partial {x_i}\partial {x_j}}}} \right)}^2}} } \\
 <& {Q^2}\sum\nolimits_{i = 1}^N {{{\left| {{o_i}} \right|}^2}} \left( {{{\left( {\sum\nolimits_{i = 1}^N {\left| {{o_i}} \right|} } \right)}^2} + \sum\nolimits_{i = 1}^N {{{\left| {{o_i}} \right|}^2}} } \right) \buildrel \Delta \over = Z.
\end{split}
\end{equation}
Since
  \begin{equation} \nonumber
 {\left\| {\nabla f_2^2\left( {{\bf{B}}\left( {{\bf{x}},\left\{ {{\theta _m}} \right\}_{m = 1}^M} \right)} \right)} \right\|_2}{{\bf{I}}_N}\underline  \succ \nabla f_2^2\left( {{\bf{B}}\left( {{\bf{x}},\left\{ {{\theta _m}} \right\}_{m = 1}^M} \right)} \right),
 \end{equation}
  we can chose $\delta $ as $\delta  = \sqrt Z $ to strictly satisfy $\delta {{\bf{I}}_N}\underline  \succ  {\nabla ^2}{f_2}\left( {{\bf{B}}\left( {{\bf{x}},\left\{ {{\theta _m}} \right\}_{m = 1}^M} \right)} \right)$.

  \end{appendices}

\end{document}